\documentclass[pre,twocolumn,twoside,showpacs,floatfix]{revtex4}
\usepackage{dcolumn}%
\usepackage{bm}%
\usepackage{graphicx,textcomp}
\usepackage{amssymb,amsfonts,amsmath}

\begin{document}

\title{Exact combinatorial approach to finite coagulating systems}

\author{Agata Fronczak, Anna Chmiel, Piotr Fronczak}

\affiliation{Faculty of Physics, Warsaw University of Technology,Koszykowa 75, PL-00-662 Warsaw, Poland}

\date{\today}


\begin{abstract}
The paper outlines an exact combinatorial approach to finite coagulating systems. In this approach, cluster sizes and time are discrete, and the binary aggregation alone governs the time evolution of the systems. By considering the growth histories of all possible clusters, the exact expression is derived for the probability of a coagulating system with an arbitrary kernel being found in a given cluster configuration when monodisperse initial conditions are applied. Then, this probability is used to calculate the time-dependent distribution for the number of clusters of a given size, the average number of such clusters and that average's standard deviation. The correctness of our general expressions is proved based on the (analytical and numerical) results obtained for systems with the constant kernel. In addition, the results obtained are compared with the results arising from the solutions to the mean-field Smoluchowski coagulation equation, indicating its weak points. The paper closes with a brief discussion on the extensibility to other systems of the approach presented herein, emphasizing the issue of arbitrary initial conditions. 
\end{abstract}


\maketitle

\section{Introduction and a brief state-of-the-art}\label{SectIntro}

The simplest example of the coagulation process is the evolution of a closed system of clusters that join irreversibly during binary collisions (so-called coagulation acts), according to the following scheme:
\begin{equation}
(g)+(l)\stackrel{K(g,l)}{\longrightarrow}(g+l),
\end{equation}
where $(g)$ stands for a cluster of mass $g$ and $K(g,l)$ is the coagulation kernel representing the rate of the process. Over time, the number of clusters in the system decreases, and eventually all clusters merge into a single cluster.   

Coagulation, which is also called aggregation, coalescence, gelation, etc., is ubiquitous in nature. It underlies many phenomena we know in everyday life, including milk curdling, blood coagulating, clouds and smog forming, and even traffic jamming up. The phenomena mentioned above and similar ones are  the basis for certain technological applications in food processing, water treatment, clinical diagnostics, and road monitoring systems, and aggregation is also of great interest in pure sciences, including physics \cite{2010_bookKrapivsky, 2006_PhysDWattis, 2003_PhysRepLeyvraz}, chemistry \cite{1987_bookSonntag, 1984_bookFamily, 1972_bookDrake}, biology \cite{2005_bookHein}, and mathematics \cite{2006_bookBertoin, 2006_bookPitman, 1999_RevAldous}, because it \textit{beautifully illustrates some paradigmatic features of non-equilibrium phenomena, such as scaling, phase transitions, and non-trivial steady states} (see \cite{2010_bookKrapivsky}, p.~133).

There are many approaches to modeling coagulation. The best-known approach relies on the famous Smoluchowski coagulation equation \cite{1916_Smoluchowski}, which constitutes an infinite system of coupled nonlinear differential equations and provides mean-field time evolution of the cluster size distribution. Explicit, analytical solutions for Smoluchowski's equation are known only for some particular kernels (e.g., constant ($K(g,l)=const$), additive ($K(g,l)=g+l$), and multiplicative ($K(g,l)=gl$) and for selected initial conditions (e.g., monodisperse initial conditions, under which all clusters are the same size). However, considerable literature exists on the existence and uniqueness of solutions to some general classes of discrete and continuous kernels (herein, the terms \textit{discrete} and \textit{continuous} refer to the possible values taken by cluster sizes) (see, for example,~\cite{1980_White, 1980_Ziff, 1983_Hendriks, 1986_Dongen, 1994_Kreer, 2006_PhysDLeyvraz, 2015_JChemPhysBurnett}). For instance, it has been shown that for homogeneous kernels, which satisfy $K(\alpha g,\alpha l)=\alpha^\gamma K(g,l)$, the large-time behavior of solutions for Smoluchowski's equation falls into different universality classes \cite{2004_Menon}, known as self-similar dynamical scaling solutions, which depend on the characteristic exponent $\gamma$ and on the initial conditions.     

Despite the great importance of Smoluchowski's equation, it has three serious weaknesses. First, it does not cope well with  so-called gelling kernels, an example of which is the multiplicative kernel, in which case, an attempt to interpret the exact solution leads to a surprising conclusion that the total mass concentration in the coagulating system ceases to conserve after a finite time $t_c$. This occurs simultaneously with the divergence of the second moment of the cluster size distribution. Today, it is well understood that the mass deficiency is a sign of the sol-gel transition, which is attributed to the emergence of an infinite cluster (a gel). Nevertheless, it is remarkable that the sol-gel transition does not directly follow from Smoluchowski's equation. It is, in a sense, analyzed collaterally and appears only to restore the mass conservation. The second weakness is that this equation is \textit{scholastically incomplete}, describing only the average behavior of coalescing clusters and ignoring deviations from it. Finally, the equation provides a kind of infinite-volume solution for the coagulation process, due to the fact that solutions to the equation are normalized with respect to the initial condition, and therefore they expire when the system moves away from the initial state.

With respect to these shortcomings, many questions arise. For example, how big must a system be so that Smoluchowski's equation correctly describes its behavior, especially in the limit of large times? This question has been posed by previous research~\cite{1968_Marcus, 1974_Bayewitz, 1978_Lushnikov, 1985_Hendriks}, which proposed the basis of a new stochastic approach to finite coagulating systems, as opposed to deterministic, mean-field, and infinite-volume approach that dates to Smoluchowski \cite{1916_Smoluchowski}. Today, one could add many more questions. For example, can a gel phase be observed in a system with a constant kernel and initial conditions, which according to the dynamical scaling solutions of Smoluchowski's equation \cite{2004_Menon}, lead to a mass deficiency? What about other kernels that are considered non-gelling, e.g. the additive kernel? It seems that Smoluchowski's equation is not well suited to studying these problems. Therefore, a better perspective is provided by the above-mentioned stochastic approach, which has been considerably developed by Lushnikov over the last dozen or so years (see, for example, the review paper \cite{2006_PhysDLushnikov}). 

Lushnikov's contribution was related to the not-at-all-obvious observation that the master equation governing the time evolution of the probability distribution over possible states of the coagulating system, when reduced to an equation for the generating functional of this distribution, acquires a similarity to Schr\"{o}dinger's equation for interacting quantum Bose fields. This observation enabled Lushnikov to analyze the coagulating systems with constant \cite{2011_JPhysALushnokov} and multiplicative \cite{2004_PRLLushnikov, 2005_PRELushnikov} kernels, both of which began their evolution from monodisperse initial conditions. 

This paper addresses finite coagulating systems, just as did Lushnikov et al. \cite{1968_Marcus, 1974_Bayewitz, 1978_Lushnikov, 1985_Hendriks, 2006_PhysDLushnikov, 2011_JPhysALushnokov, 2004_PRLLushnikov, 2005_PRELushnikov}.  However, in the approach described herein, unlike in the work of our predecessors, time is discrete; therefore, we begin not with the master equation, but by assuming that a single coagulation act occurs in each time step. For successive steps, we define the space of available states, and then, by studying the growth histories of all clusters, we determine the probability distribution over that space. 

The paper is organized as follows. Section~\ref{sectII} provides a thorough introduction to our method, which uses certain combinatorial structures, the so-called Bell polynomials, which are discussed in detail. This section derives the exact expression for the probability distribution that a coagulating system with an arbitrary kernel will be found in a given cluster configuration when monodisperse initial conditions are applied. In Section~\ref{sectIII}, the obtained distribution is used to calculate various cluster statistics, including the average number of clusters of a given size, its standard deviation, and the probability distribution for the number of clusters of a given size. The above-mentioned general calculations are tested for the constant kernel, providing a number of exact results that have heretofore been unknown. Section~\ref{sectIV} contains concluding remarks and briefly discusses the problem of arbitrary initial conditions, and the issue of the continuous-time, which enable direct comparison of our results with those of other approaches.  

\section{Probability distribution over the state space}\label{sectII}

We begin this section by making some simple observations about the system under investigation. First, beginning with monodisperse initial conditions, if a single coagulation act occurs in each time step, at time $t$ we have exactly
\begin{equation}\label{defk}
k=N-t
\end{equation}
clusters or particles (monomers, dimers, trimers, etc.), where $N$ is the number of monomeric units in the system. Second, the state of the system can be described as:
\begin{eqnarray}\label{defOmega}
\Omega&=&\{n_1,n_2,\dots,n_g,\dots,n_N\}
\end{eqnarray}
where $n_g\geq 0$ is the number of clusters of mass $g$, with $g$ being the number of monomeric units. Of course, in (\ref{defOmega}), the sequence $\{n_g\}$ is not arbitrary, but due to the evolution of the system, it satisfies the following equations
\begin{equation}\label{ngEq}
\sum_{g=1}^Nn_g=k,\;\;\;\mbox{and}\;\;\;\sum_{g=1}^Ng\,n_g=N.
\end{equation}
Third, the total number of states, $\overline{\Omega}$, to which the coagulation process leads, depends on time, and it is easy to deduce that it is given by the Stirling number of the second kind 
\begin{equation}\label{SNk}
\overline{\Omega}(t)=S(N,k),
\end{equation}
which describes the number of ways to partition a set of $N$ objects into $k$ subsets.

\begin{figure}[]
	\centering\includegraphics[width=0.98\columnwidth]{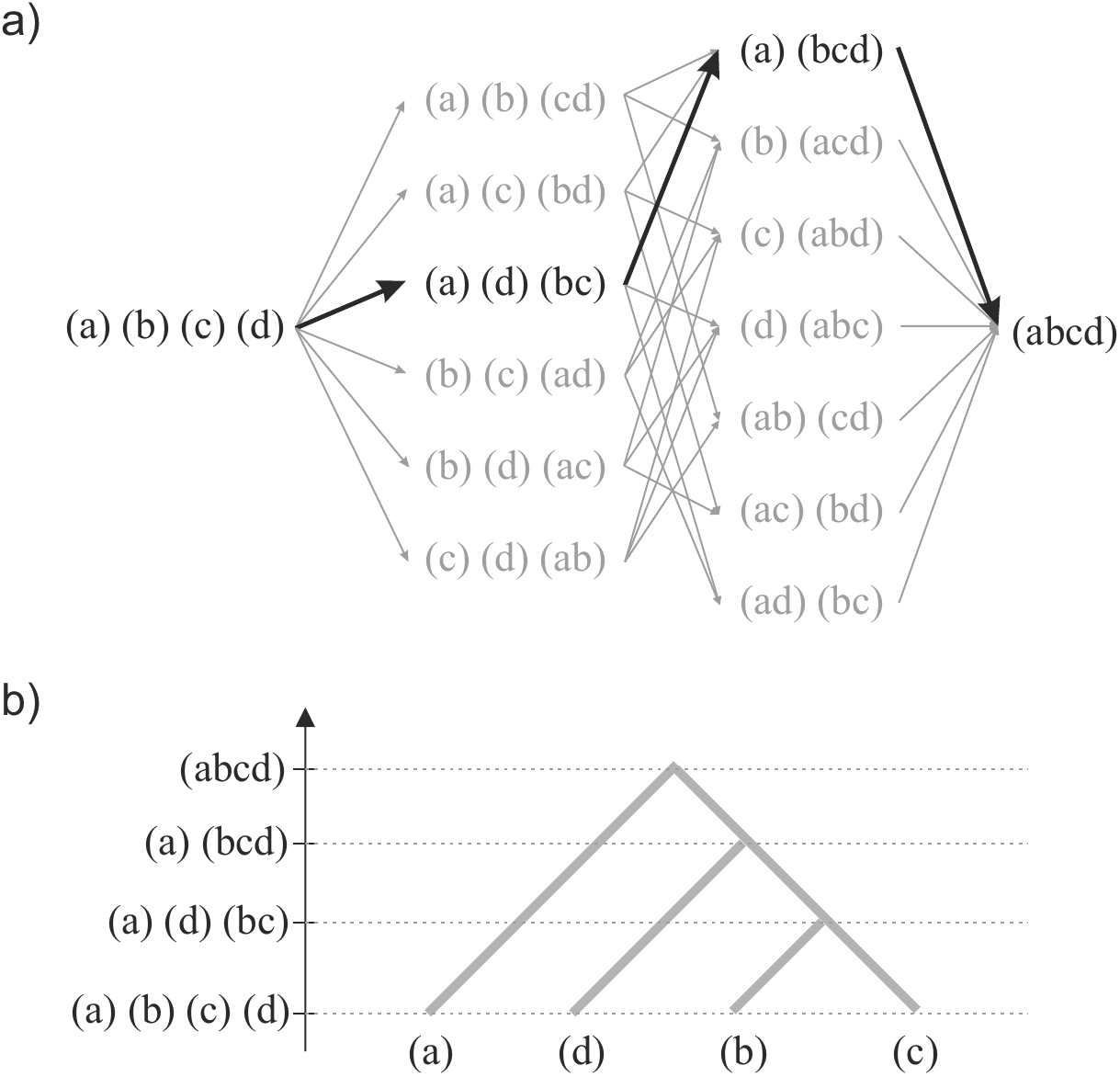}
	\caption{a) \textbf{Diagram illustrating all possible growth histories of particles in the case of the constant kernel}. The resulting particle, (abcd), consists of four $g=4$ tagged monomeric units: (a)(b)(c)(d). Its growth requires $g-1=3$ coagulation acts, which are illustrated by arrows. After the first coagulation act, the future cluster consists of two monomers and one dimer. After the second time step, it consists of two parts: either one monomer and one trimer, or two dimers. In the third step, the particle is formed. The number of different growth histories, $x_g$, is equal to the number of different paths drawn by arrows and leading through different states of the diagram. For $g=4$, the number is equal to $x_4=18$ ( cf.~Eq.~(\ref{Wxg})).
	b) \textbf{Sample tree corresponding to the bold path in the diagram}. Every particles' growth history can be illustrated as a rooted tree, with leaf nodes standing for monomeric units, internal nodes representing the history-dependent transition states of the cluster, and the root node being the last step in the cluster's growth process.}
	\label{reffig1}
\end{figure}

\begin{figure}[]
	\centering\includegraphics[width=0.99\columnwidth]{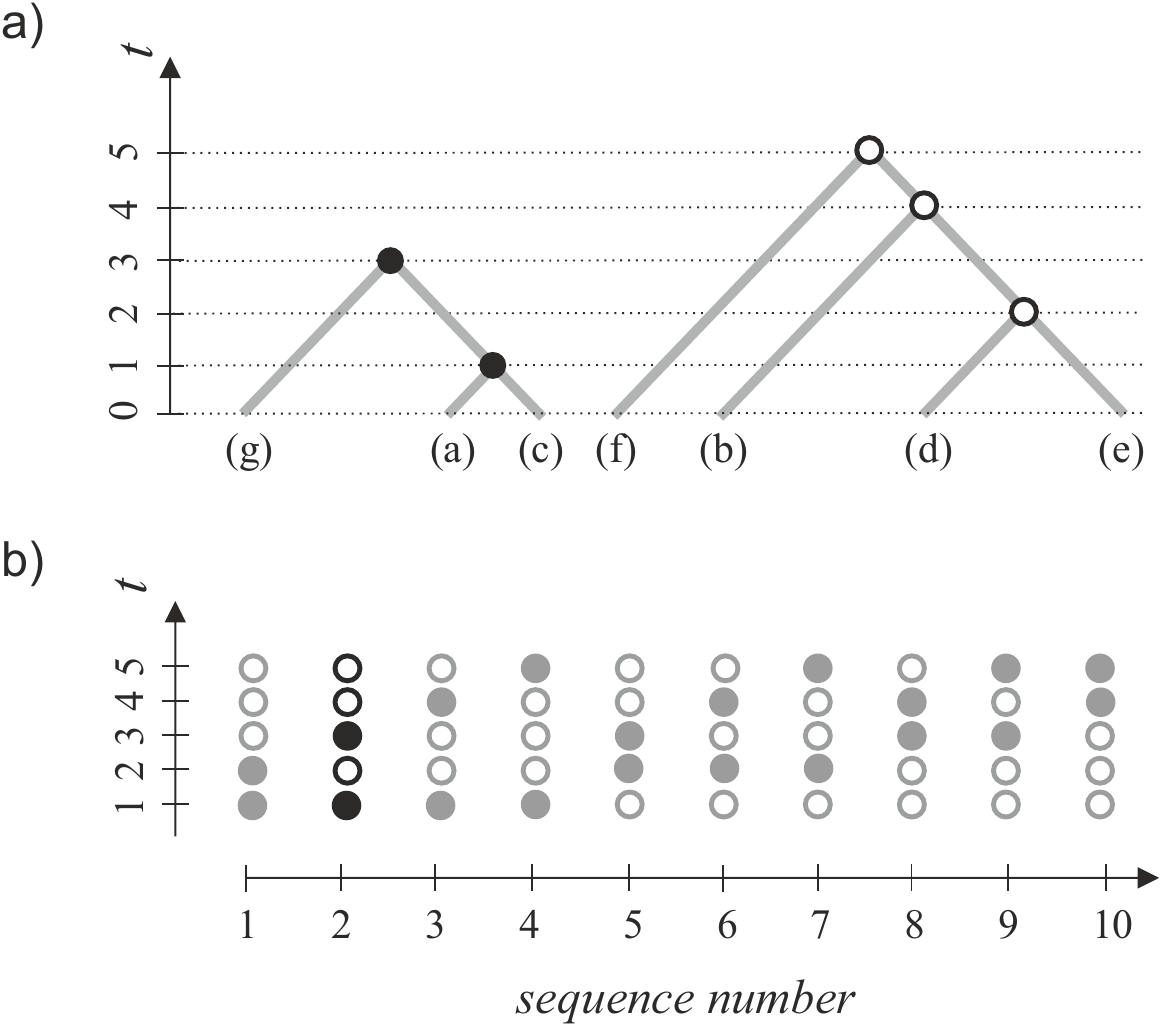}
	\caption{\textbf{Pictorial representation of the time evolution of a coalescing system with a constant kernel}. a)~One of the many possible microscopic realizations of the system of $N\!=\!7$ monomeric units at time $t\!=\!5$. The system consists of $k\!=\!2$ (cf.~Eq.(\ref{defk})), particles of sizes $3$ and~$4$. The microstate shown, (acg)(bdef), contributes to the state $\Omega=\{0,0,1,1,0,0,0\}$ (cf.~Eq.~(\ref{defOmega})), in which $n_3=n_4=1$ and all other numbers $n_g$ are equal to zero. Nodes in trees corresponding to different clusters are marked with different symbols (closed and open circles, respectively) to emphasize that a given microstate in which particles have the same history of coagulation acts (i.e. the same structure of the corresponding trees) can be created in many ways. This is so because the coagulation acts corresponding to different clusters may alternate with each other. In particular, the microstate shown in a) corresponds to the second of a total of ten sequences shown in b). The sequence can be written as (a)(b)(c)(d)(e)(f)(g) $\rightarrow$ (b)(d)(e)(f)(g)(ac) $\rightarrow$ (b)(f)(g)(ac)(de) $\rightarrow$ (b)(f)(de)(acg) $\rightarrow$ (f)(acg)(bde) $\rightarrow$ (acg)(bdef). For yet another example, one can consider the last sequence in b) which corresponds to the following arrangement of  coagulation acts: (a)(b)(c)(d)(e)(f)(g)$\rightarrow$ (a)(b)(c)(f)(g)(de) $\rightarrow$ (a)(c)(f)(g)(bde) $\rightarrow$ (a)(c)(g)(bdef) $\rightarrow$ $\rightarrow$ (g)(ac)(bdef) $\rightarrow$ (acg)(bdef). For an arbitrary state $\Omega$, the total number of such sequences is given by Eq.~(\ref{W2b}).}
	\label{figsigma}
\end{figure}

For further derivations, it is important to introduce the so-called partial (or incomplete) exponential Bell polynomials \cite{book_Comtet} (hereafter called Bell polynomials), $B_{N,k}(x_1,x_2,\dots,x_{N-k+1})=B_{N,k}(\{x_g\})$, which have a few features that make them very useful for analyzing aggregation phenomena. The polynomials are defined as
\begin{equation}\label{defBell}
B_{N,k}(\{x_g\})=N!\sum_{\{n_g\}}\prod_{g=1}^{N-k+1}\frac{1}{n_g!} \left(\frac{x_g}{g!}\right)^{n_g},
\end{equation} 
where the summation is taken over all non-negative integers $\{n_g\}$ that satisfy Eqs.~(\ref{ngEq}). It takes a moment to see that the polynomials encode very detailed information related to the ways in which an arbitrary set can be partitioned. Suppose that $N$ distinguishable objects are partitioned into $k$ non-empty and disjoint subsets of $c_i>0$ elements each, where $\sum_{i=1}^kc_i=N$. There are exactly
\begin{equation}\label{Bell1}
\binom{N}{c_1,c_2,\dots,c_k}=N!\prod_{i=1}^k\frac{1}{c_i!}=N!\prod_{g=1}^{N-k+1}\left(\frac{1}{g!}\right)^{n_g}
\end{equation}
of such partitions, where $n_g\geq 0$ stands for the number of subsets of size $g$, with the largest subset size being equal to $N-k+1$. Further suppose that in such a composition, each of $n_g$ subsets of size $g$ can be in any of $x_g\geq 0$ internal states and that the order of clusters does not matter. Then, the number of partitions becomes
\begin{equation}\label{Bell2}
N!\prod_{g=1}^{N-k+1}\frac{1}{n_g!}\left(\frac{x_g}{g!}\right)^{n_g}.
\end{equation}
Summing Eq.~(\ref{Bell2}) over all integers $\{n_g\}$ specified by Eq.~(\ref{ngEq}) one obtains the partial Bell polynomial $B_{N,k}(\{x_g\})$, which is defined by Eq.~(\ref{defBell}). From the above explanations, it is easy to realize that the Stirling partition number, $S(N,k)$ (Eq.~(\ref{SNk})), is simply the value of the Bell polynomial $B_{N,k}(\{x_i\})$ on the sequence of ones: $S(N,k)=B_{N,k}(1,1,\dots,1)$.

For example, if we consider a set of $N=3$ monomers (a)(b)(c), the set can be partitioned into $k=2$ clusters in three ways: (a)(bc), (b)(ac), and (c)(ab). This partitioning is described by the corresponding Bell polynomial as follows: $B_{3,2}({x_1,x_2})=3x_1x_2$. Similarly, in the case of $N=6$ monomeric units and $k=3$ particles one would obtain: $B_{6,3}(x_1,x_2,x_3,x_4)=15x_1^2x_4+60x_1x_2x_3+15x_2^3$, because there are
$15$ ways to partition a set of $6$ as $1+1+4$, $60$ ways to partition such a set as $1+2+3$, and $15$ ways to partition it as $2+2+2$. Accordingly, in the two examples above, one obtains: $S(3,2)=3$ and $S(6,3)=90$. 

Now, after introducing the general concept of aggregation and acquainting readers with the necessary definitions, our aim is to derive the probability, $P(\Omega)$, of a coagulating system being found in a given state $\Omega$ (Eq.~(\ref{defOmega})). Due to the non-equilibrium characteristic of the process investigated, at time $t$, the allowed states of the system are not equiprobable, that is $P(\Omega)\neq\overline{\Omega}(t)^{-1}$. To find the probability distribution function $P(\Omega)$ over the time-dependent state space $\{\Omega\}$, one must determine the thermodynamic probabilities, $W(\Omega)$, which stand for the number of ways in which the corresponding state $\Omega$ can be obtained as a result of the time evolution of the system. Knowing thermodynamic probabilities, one would immediately have: $P(\Omega)=W(\Omega)Z^{-1}$, where $Z=\sum_{\Omega}W(\Omega)$.

Fortunately, both $W(\Omega)$ and $Z$ can be found easily with the help of methodology that is covered by the Bell polynomials. The starting point for our reasoning is Eq.~(\ref{Bell2}), which describes the number of ways in which $N$ monomers can be partitioned into $k$ clusters. However, there are adjustments that must be made. First, the number $x_g$, which characterize the internal states of a single cluster of size $g$ should be equal to the number of ways in which the cluster can be created from tagged monomeric units. Obviously, $x_g$ must depend on the number of monomers, $g$, and on the method of combining them into the particle, that is, on the kernel used (for an illustrative example of the constant kernel, see Fig.~1). Second, when applied directly, Eq.~(\ref{Bell2}) tacitly assumes that all clusters arise at once, is other words, at the same time step. Of course, this is not true. A single cluster of size $g$ arises as a result of $g-1$ coagulation acts. Furthermore, the acts corresponding to different clusters may alternate with each other. The above gives rise to a multiplication effect in the number of ways a given microstate can be created (see Fig. 2). In the following, we discuss these two issues quantitatively.

Let us start with $x_g$, which is the number of ways in which a cluster of size $g$ can be created from a given subset of tagged monomeric units. In the case of the constant kernel, it can be written as 
\begin{equation}\label{Wxg}
x_g=\binom{g}{2}\binom{g\!-\!1}{2}\binom{g\!-\!2}{2}\dots\binom{2}{2}= \frac{g!(g\!-\!1)!}{2^{g\!-\!1}}.
\end{equation}
The above expression simply states the following. In the first time step, one chooses and coalesces two clusters (i.e. monomers) from the $g$ available. In the second time step, one has $g-1$ clusters (i.e. $g-2$ monomers and one dimer, correspondingly) two of which are chosen and merged. In the third step, one selects the next two clusters available out of $g-2$, and so on.

Now, having the sequence $\{x_g\}$ (Eq.~(\ref{Wxg})), one can use Eq.~(\ref{Bell2}) to calculate the number of different partitions of tagged monomers into a given set of clusters $\{n_g\}$, in which every cluster's evolution is considered. This number is not yet equal to $W(\Omega)$, due to the fact that although the individual evolution of every cluster is covered by the sequence $\{x_g\}$ the global inter-cluster time evolution is not yet  taken into account. To be precise, a given state can be obtained as a result of different sequences of intermixed coagulation acts corresponding to different clusters. As already mentioned each particle of size $g$ requires $g-1$ coagulation acts in order to be created. Thus, since the total number of coagulation acts is equal to
\begin{equation}\label{t}
\sum_{g=1}^{N-k+1}(g-1)n_g=N-k=t,
\end{equation}
it is easy to deduce that the overall number of such sequences corresponding to each of (\ref{Bell2}) microscopic realizations of the system is equal to
\begin{widetext}
\begin{eqnarray}
\label{W2a}\left[\binom{t}{1}\!\binom{t\!-\!1}{1}\!\dots\!\binom{t\!-\!n_2\!+\!1}{1}\right]\left[\binom{t\!-\!n_2}{2}\!\binom{t\!-\!n_2\!-\!2}{2}\!\dots\!
\binom{t\!-\!n_2\!-\!2(n_3\!-\!1)}{2}\right]&\dots&
\\\label{W2b}=\frac{t!}{(1!)^{n_2}(2!)^{n_3}\dots((g-1)!)^{n_g}\dots}
&=&t!\prod_{g=2}^{N-k+1}\frac{1}{((g-1)!)^{n_g}}\;.
\end{eqnarray}
\end{widetext}
In the above expression, the consecutive square brackets refer to dimers, trimers, etc. In the brackets, the product of binomial coefficients states the number of ways in which the $g-1$ coagulation acts corresponding to successive $g$-mers can be deployed in the timeline. To further clarify, let us note that the number of coagulating acts corresponding to monomers is $n_g(g-1)=0$, where $g=1$. Therefore, in Eqs.~(\ref{W2a})-(\ref{W2b}), one begins with dimers, each of which requires one connection act. Then we have trimers, with two coagulation acts each, and so on. 

Finally, by multiplying~(\ref{Bell2}) and~(\ref{W2b}), one gets the exact formula for the thermodynamic probability, $W(\Omega)$, which is the number of ways in which the state $\Omega$ can be obtained
\begin{widetext}
\begin{equation}\label{WOmega}
W(\Omega)=\left[t!\prod_{g=1}^{N-k+1}\frac{1}{((g-1)!)^{n_g}}\right]
\left[N!\prod_{g=1}^{N-k+1}\frac{1}{n_g!} \left(\frac{x_g}{g!}\right)^{n_g}\right]=
t!\,N!\prod_{g=1}^{N-k+1} \frac{1}{n_g!}\left(\frac{x_g}{(g-1)!g!}\right)^{\!n_g}.
\end{equation}
Accordingly, with the help of the Bell polynomials, the sum of $W(\Omega)$ over all the systems' states can be calculated
\begin{equation}\label{Za}
Z=\sum_{\Omega}W(\Omega)=t!\left[N!\sum_{\{n_g\}}
\prod_{g=1}^{N-k+1}\frac{1}{n_g!}\left(\frac{x_g}{(g-1)!g!}\right)^{\!n_g} \right]\;\stackrel{Eq.(\ref{defBell})}{=}\;t!\,B_{N,k}\left(\left\{\frac{x_g} {(g-1)!}\right\}\right)
\;\stackrel{Eq.(\ref{defyg})}{=}\;t!\,B_{N,k}(\{y_g\}),
\end{equation}
\end{widetext}
where, in order to simplify the calculations below, a new parameter is introduced:
\begin{equation}\label{defyg}
y_g=\frac{x_g}{(g-1)!}.
\end{equation}

Now, we would like to comment on Eqs.~(\ref{WOmega}) and~(\ref{Za}) which are the most important results of this paper. They exactly specify the probability distribution, 
\begin{eqnarray}\label{defPO}
P(\Omega)\!=\!\frac{W(\Omega)}{Z}\!=\!
\frac{N!}{B_{N,k}(\{y_g\})}\!\prod_{g=1}^{N-k+1}\!\frac{1}{n_g!}\left(\frac{y_g}{g!}\right)^{\!n_g}\!,
\end{eqnarray}
for a coagulating system being found in a given state $\Omega$ when monodisperse initial conditions are applied. The only place where the kernel information is encoded is the sequence $\{x_g\}$. Strictly speaking, Eq.~(\ref{defPO}) provides the most detailed information about the finite-size coalescing system, which has not previously been known. The distribution obtained over the time-dependent state space is the equivalent of the Boltzmann distribution which is inapplicable to non-equilibrium systems (like those we study) due to its insensitivity to the direction of time. 

Correspondingly, in the case of the constant kernel, the obtained expressions can be rewritten as follows:
\begin{equation}\label{WOmegaa}
W(\Omega)\stackrel{Eq.(\ref{Wxg})}{=}\; t!\,N!\prod_{g=1}^{N-k+1}\frac{1}{n_g!}\frac{1}{2^{(g-1)n_g}}\;,
\end{equation}
and 
\begin{align}\label{Z1a}
Z\;\stackrel{Eq.(\ref{pom1Bell})}{=}\;&t!\,
\frac{2^k}{2^N}B_{N,k}(\{g!\})\\\label{Z1b}
\;\stackrel{Eq.(\ref{pom2Bell})}{=}\;&\frac{1}{2^t}\frac{N!}{(N\!-\!t)!}
\frac{(N\!-\!1)!}{(N\!-\!1\!-\!t)!}\;,
\end{align}
where the identity $k=N-t$~(\ref{defk}) and some basic properties of the Bell polynomials \cite{book_Comtet} have been used:
\begin{align}\label{pom1Bell}
\mbox{i.}\;\;B_{N,k}&(\{ab^{g}x_g\})=a^kb^NB_{N,k}(\{x_g\}),\hspace{1.5cm} \\\label{pom2Bell}
\mbox{ii.}\;\;B_{N,k}&(\{g!\})=\binom{N\!-\!1}{k\!-\!1}\frac{N!}{k!}. \hspace{1.5cm}
\end{align}
Finally, for the constant kernel, the probability distribution, Eq.~(\ref{defPO}), becomes
\begin{equation}\label{defPOa}
P(\Omega)=\frac{2^t(N\!-\!t)!}{\binom{N-1-t}{t}}
\prod_{g=1}^{N-k+1}\frac{1}{n_g!}\frac{1}{2^{(g-1)n_g}}.
\end{equation}

In the following section, we use Eqs.~(\ref{defPO}) and~(\ref{defPOa}) to derive time-dependent cluster statistics in finite coagulating systems. 

\section{Cluster statistics}\label{sectIII}

\subsection{Average number of clusters of a given size \\and the standard deviation of the average}

Once the probability distribution over the state space of the coagulating system, Eq.~(\ref{defPO}), is determined, one can proceed to calculate the average number of clusters of a given size and the standard deviation of the average. For these calculations we use the expression for the derivative of the Bell polynomials \cite{FaaDiBruno}: 
\begin{equation}\label{pom3aBell}
\frac{\partial B_{N,k}(\{x_g\})}{\partial x_s} =\binom{N}{s}B_{N-s,k-1}(\{x_g\})\;\;\;\mbox{for}\;\;\;s\!\in\!A
\end{equation}
 and
\begin{equation}\label{pom3bBell}
\frac{\partial B_{N,k}(\{x_g\})}{\partial x_s} =0 \;\;\;\;\mbox{for}\;\;\;\;s\!\in S\backslash A, 
\end{equation}
where 
\begin{equation}\label{defAS}
A\!=\!\{1,\dots,N\!-\!k\!+\!1\}\;\;\;\;\mbox{and}\;\;\;\;
S\!=\!\{1,2,\dots,N\}.
\end{equation}

\begin{figure}[]
	\centering\includegraphics[width=0.90\columnwidth]{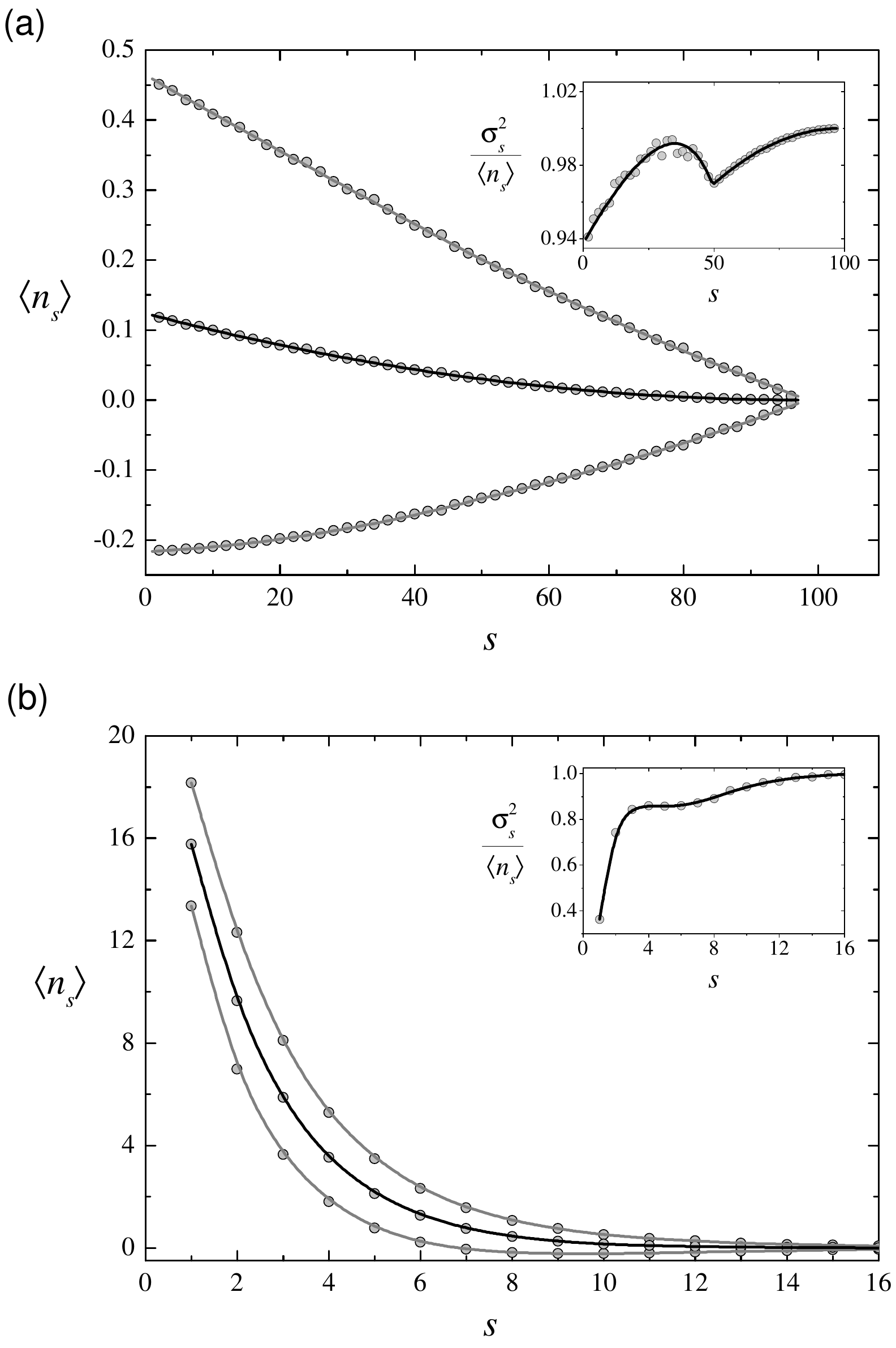}
	\caption{\textbf{Cluster statistics in coagulating systems with a constant kernel} arising from monodisperse initial conditions with $N=100$ monomeric units, and (a) $k=4$ or (b) $k=40$ clusters, respectively. Main panels: mean number of clusters of a given size and its standard deviation. Solid lines correspond to theoretical predictions: black lines for $\langle n_s\rangle$, Eqs.~(\ref{nsK1})-(\ref{nsK2}), and grey lines for $\langle n_s\rangle\pm \sigma_s$ (Eqs.~(\ref{sdK1})-(\ref{sdK2})). The scattered points represent the results of numerical simulations averaged over $10^5$ independent realizations of the model. Insets: Variance divided by the mean.}
	\label{reffig3}
\end{figure}

Thus, in the system with $N$ monomeric units and $k$ clusters, the expression for the average number of clusters of size $s$ can be calculated as follows:
\begin{align}\label{ns1}
\langle n_s\rangle&=\sum_{\Omega}n_s(\Omega)P(\Omega)\\\label{ns2}&= \frac{N!}{B_{N,k}(\{y_g\})}\sum_{\{n_g\}}n_s\prod_g\frac{1}{n_g!}
\left(\frac{y_g}{g!}\right) ^{\!n_g}\!\\\label{ns3}&=
\frac{N!}{B_{N,k}(\{y_g\})}\sum_{\{n_g\}}
\left(y_s\frac{\partial}{\partial y_s}\right)\prod_g \frac{1}{n_g!} \left(\frac{y_g}{g!}\right)^{\!n_g}\\\label{ns4}&=
\frac{1}{B_{N,k}(\{y_g\})}\left(y_s\frac{\partial}{\partial y_s}\right) B_{N,k}(\{y_g\})\\\label{ns5}&\stackrel{Eq.(\ref{pom3aBell})}{=}
\binom{N}{s}y_s\frac{B_{N-s,k-1}(\{y_g\})}{B_{N,k}(\{y_g\})} \;\;\;\mbox{for}\;\;\;s\!\in\!A, 
\end{align}
and
\begin{equation}\label{ns6}
\langle n_s\rangle=0\;\;\;\mbox{for}\;\;\;s\!\in\!S\backslash A.
\end{equation}
Correspondingly, the standard deviation of this average is given by:
\begin{align}
\sigma_s&=\sqrt{\langle n_s^2\rangle-\langle n_s\rangle^2}\\ 
&=\sqrt{\langle n_s(n_s\!-\!1)\rangle+\langle n_s\rangle-\langle n_s\rangle^2},
\end{align}
where
\begin{equation}\label{sd1}
\langle n_s(n_s\!-\!1)\rangle=\binom{N}{s,\!s}y_s^2
\frac{B_{N-2s,k-2}(\{y_g\})}{B_{N,k}(\{y_g\})}\;\;\mbox{for}\;\;s\!\in\!B,
\end{equation}
and 
\begin{equation}\label{sd2}
\langle n_s(n_s\!-\!1)\rangle=0\;\;\;\mbox{for}\;\;\;s\!\in\!S\backslash B,
\end{equation}
with 
\begin{equation}\label{defB}
B=\{1,\dots,(N\!-\!k)/2\!+\!1\}.
\end{equation}

\begin{figure}[]
	\centering\includegraphics[width=0.99\columnwidth]{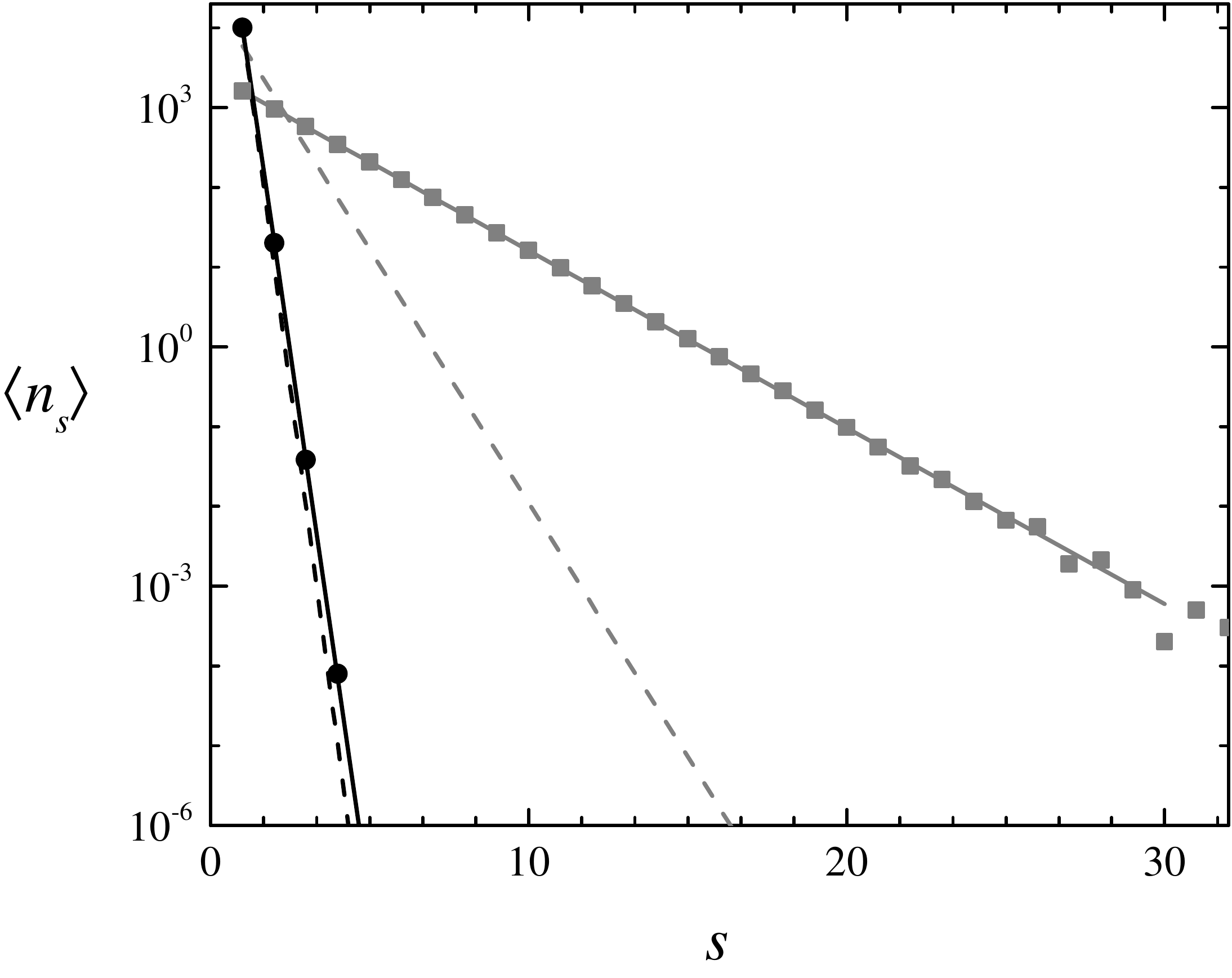}
	\caption{\textbf{Comparison of our approach with the results obtained from Smoluchowski's equation}. Solid lines represent our combinatorial expressions (Eqs.~(\ref{nsK1})-(\ref{nsK2})). Dashed lines represent the exact solution of the discrete version of  Smoluchowski's equation (Eq.~(\ref{Smoluchowski})). Scattered points represent the results of the numerical simulations of coagulating systems of size $N=10^4$ and $k=4000$ (gray squares) or $k=9980$ (black circles), averaged over $10^{6}$ independent realizations of the model.}
	\label{reffig4}
\end{figure}

In the case of the constant kernel, when
\begin{equation}\label{defyg0}
y_g=\frac{g!}{2^{g-1}},
\end{equation} 
(cf.~Eqs.~(\ref{Wxg}) and~(\ref{defyg})), the above expressions for the expected value and the standard deviation of the number of clusters of a given size simplify to:
\begin{align}\label{nsK1}
\langle n_s\rangle&=k\frac{\binom{N-1-s}{k-2}}{\binom{N-1}{k-1}} 
\;\;\;\mbox{for}\;\;s\!\in\!A,\\\label{nsK2}
\langle n_s\rangle&=0\;\;\;\;\;\;\;\;\;\;\;\;\;\;\;\;\; \mbox{for}\;\;s\!\in\!S\backslash A,
\end{align}
and
\begin{align}\label{sdK1}
\langle n_s(n_s\!-\!1)\rangle &=k(k\!-\!1)\frac{\binom{N-1-2s}{k-3}}{\binom{N-1}{k-1}} 
\;\;\;\mbox{for}\;\;s\!\in\!B,\\\label{sdK2}
\langle n_s(n_s\!-\!1)\rangle&=0\;\;\;\;\;\;\;\;\;\;\;\;\;\; \;\;\;\;\;\;\;\;\;\;\;\;\;\mbox{ for}\;\;s\!\in\!S\backslash B.
\end{align}

Figures.~\ref{reffig3} and~\ref{reffig4} show excellent agreement between our theoretical predictions and the results of the numerical simulations performed for coalescing systems with a constant kernel and arising from monodisperse initial conditions. As the numerical simulations show, the agreement is independent of the parameters of the model. Even
for small values of the system size, like N = 100 (Fig. 3a), our theoretical predictions perfectly reproduce not only
the mean number of clusters of a given size and its standard deviation, but also non-monotonic relation between
the variance and the mean.

Figure~\ref{reffig4} also shows that the exact solution of the discrete version of Smoluchowski's equation (see Table~2 in~\cite{1999_RevAldous}),
\begin{equation}\label{Smoluchowski}
n_s(t)=\frac{4}{t(t\!+\!2)}\left(\frac{t}{t\!+\!2}\right)^s,
\end{equation}
agrees with the numerical simulations only in the limit of small times, $t\ll N$, that is, when the total number of clusters is comparable to the initial number of monomers, $k\leq N$, which is assumed to be very large, $N\gg 1$. This limitation does not apply to our theoretical predictions, which are in compliance with the numerical simulations also for larger value of $t$.
 
\subsection{Probability distribution for the number 
	\\of clusters of a given size}

Using Eq.~(\ref{defPO}), one can also derive the time-dependent probability distribution for the number of clusters of a given size. The first two moments of this distribution have already been calculated (see Eqs.~(\ref{ns5})-(\ref{ns6}) and~(\ref{sd1})-(\ref{sd2})). To perform this derivation concisely, we must introduce some new definitions. We also use some additional properties of the Bell polynomials. These definitions and properties will be introduced at the appropriate time, as needed. 

In what follows, we will focus on clusters of size $s$. The goal is to find $P(n_s)$, that is the probability that there are exactly $n_s$ clusters of size $s$ in the system consisting of $N$ monomers in which there are $k$ clusters in total. This probability is simply the sum:
\begin{equation}\label{defPns1}
P(n_s)=\sum_{\Omega^*}P(\Omega^*),
\end{equation} 
where the summation runs over all states $\Omega^*$ of the system, in which $n_s$ is fixed. Such states can be defined as  follows (cf.~Eqs.~(\ref{defOmega}) and~(\ref{ngEq})):
\begin{eqnarray}\label{defOmega1}
\Omega^{*}\!=\!\{n_{g}\!:\hspace{0.03cm}n_s\!=\!const\hspace{0.03cm}\wedge \hspace{0.03cm}\sum_{g\neq s}\!n_g\!=\!k^*\hspace{0.03cm}\wedge\hspace{0.03cm}
\sum_{g\neq s}\!g\hspace{0.02cm}n_g\!=\!N^*\!\},
\end{eqnarray}
where
\begin{equation}\label{kstar}
k^*=k\!-\!n_s\;\;\;\;\;\mbox{and}\;\;\;\;\;N^*=N\!-\!s\hspace{0.02cm}n_s.
\end{equation}

After inserting Eq.~(\ref{defPO}) into~(\ref{defPns1}), one obtains the following general expression for the probability distribution of the number $n_s$ of clusters of size $s$:
\begin{align}\label{defPns2}
P(n_s)\!&=\!\frac{N!}{B_{N,k}(\{y_g\})}
\frac{1}{n_s!}\left(\frac{y_s}{s!}\right)^{\!n_s} 
\sum_{\Omega^*}\!\prod_{g\neq s}\!\frac{1}{n_g!}\left(\frac{y_g}{g!}\right)^{\!n_g}\!\\\label{defPns3} \!&\stackrel{Eq.(\ref{defBell})}{=} \frac{1}{n_s!}\left(\frac{y_s}{s!}\right)^{\!n_s} \frac{N!}{N^*!}\frac{B_{N^*\!,k^*}(\{(1-\delta_{gs})y_g\})}{B_{N,k}(\{y_g\})},
\end{align}
where $\delta_{gs}$ is the Kronecker delta, and the corresponding sequence of parameters $\{y_g(1-\delta_{gs})\}$ stands for
$\{y_1,\dots,y_{s-1},0,y_{s+1},\dots,y_{N}\}$.

For the constant kernel, Eq.~(\ref{defPns3}) can be further simplified. However, before doing this, we would like to point out that the result obtained fits nicely into the longstanding research on coagulation systems. Namely, there has been a great deal of discussion of whether or not $P(n_s)$ obeys Poisson statistics (see, for example,~\cite{1974_Bayewitz}). Given numerical arguments, one conjectured that \textit{as time increases} the distribution approaches a Poisson distribution. According to our knowledge, Eq.~(\ref{defPns3}) is the first theoretical confirmation of this behavior. From this expression one immediately sees that the Poisson-like behavior is recovered when $N^*\rightarrow N$ and $k^*\rightarrow k$, which does not necessarily (although it may) agree with the phrase above in italics.

To simplify Eq.~(\ref{defPns3}) for the constant kernel, we deal separately with the Bell polynomials in the numerator and the denominator of this equation. Thus, using the previously introduced properties of these polynomials, the polynomial in the denominator can be represented as:
\begin{align}\label{PBell1a}
B_{N,k}(\{y_g\})&\stackrel{Eq.(\ref{defyg0})}{=} B_{N,k}\left(\left\{\frac{g!}{2^{g-1}}\right\}\right)\\\label{PBell1b}& \stackrel{Eq.(\ref{pom1Bell})}{=}2^{k-N}B_{N,k}(\{g!\})\\ \label{PBell1c}&
\stackrel{Eq.(\ref{pom2Bell})}{=}2^{k-N}\frac{N!}{k!}\binom{N-1}{k-1}.
\end{align} 

Accordingly, the polynomial in the numerator can be transformed as follows:
\begin{align}\label{PBell2a}
&B_{N^*,k^*}(\{(1-\delta_{gs})y_g\})=
\\&\label{PBell2b}\stackrel{Eq.(\ref{pom3Bell})}{=} 
\sum_{\nu\leq N^*}\sum_{\kappa\leq k^*}
\binom{N^*}{\nu}B_{\nu,\kappa}(\{\!-\delta_{gs}y_g\}) B_{N^*\!-\!\nu,k^*\!-\!\kappa}(\{y_g\})
\\&\label{PBell2c}\stackrel{Eq.(\ref{pom4bBell})}{=}\! \sum_{\kappa=0}^{\kappa_{max}}\!
\binom{N^*}{s\kappa}\! \left[\frac{(s\kappa)!}{\kappa!(s!)^\kappa}(-y_s)^\kappa\right]\!
B_{N^*\!-\!s\kappa,k^*\!-\!\kappa}(\{y_g\})
\\&\label{PBell2d}\stackrel{Eq.(\ref{defyg0})}{=}2^{k^*\!-\!N^*}\frac{N^*!}{k^*!} \sum_{\kappa=0}^{\kappa_{max}}\!\binom{k^*}{\kappa} \binom{N^*\!-\!s\kappa\!-\!1}{k^*\!-\!\kappa\!-\!1}(\!-\!1)^\kappa,
\end{align} 
where 
\begin{align}
\kappa_{max}&=\mbox{min}\left\{k^*,\frac{N^*\!-\!k^*}{s\!-\!1}\right\}\\ &\stackrel{Eq.(\ref{kstar})}{=}\mbox{min}\left\{k\!-\!n_s, \frac{N\!-\!k}{s\!-\!1}\!-\!n_s\right\}
\end{align}
(the second value of $\kappa_{max}$ simply results from the condition $N^*\!-\!\nu\geq k^*\!-\!\kappa$, Eq.~(\ref{PBell2b}), where $\nu=s\kappa$, Eq.~(\ref{PBell2c})), and where the below properties of the Bell polynomials~\cite{book_Comtet} have been used:
\begin{align}\label{pom3Bell}
\mbox{iii.}\;\;B_{N,k}&(\{x_g\!+\!y_g\})=
\\\nonumber=&\sum_{\nu\leq N}\sum_{\kappa\leq k}
\binom{N}{\nu}B_{\nu,\kappa}(\{x_g\})B_{N\!-\!\nu,k\!-\!\kappa}(\{y_g\}),
\\\label{pom4aBell}
\mbox{iii.}\;\;B_{N,k}&(\{\delta_{gs}x_g\})=0,\hspace{1.67cm}\mbox{for} \;\;\;\; N\neq sk,
\\\label{pom4bBell}
B_{N,k}&(\{\delta_{gs}x_g\})=\frac{(N)!}{k!(s!)^k}x_s^k,\;\;\;\;\mbox{for} \;\;\;\; N=sk.
\end{align}

\begin{figure}[]
	\centering\includegraphics[width=0.99\columnwidth]{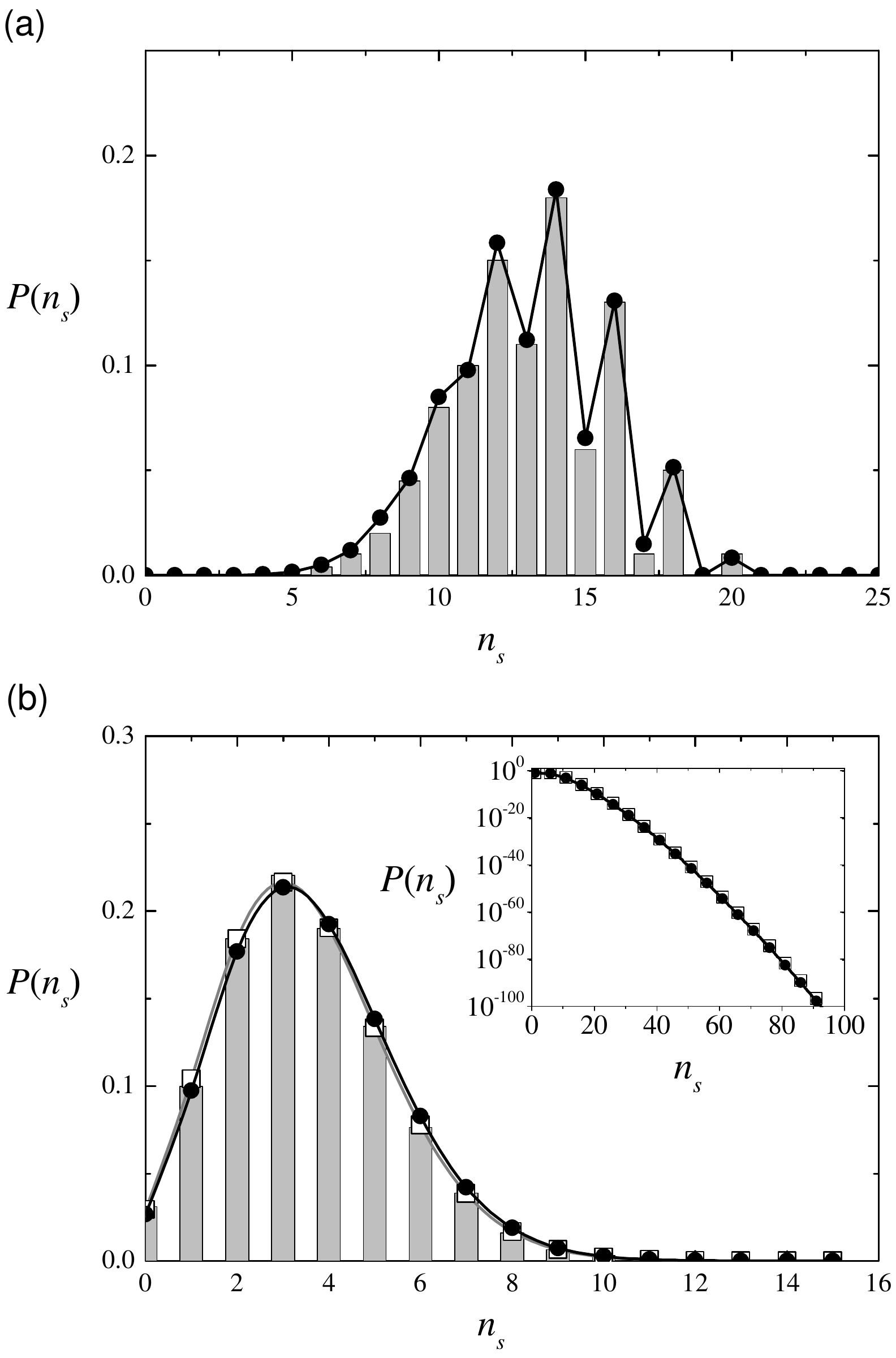}
	\caption{\textbf{Probability distributions for the number of clusters of a given size}. The following graphs show a) $N=100$, $k=80$, and $s=2$ (dimers); and b) $N=10^5$, $k=600$, and $s=5$ (clusters of size $5$). In both graphs, bars represent results of the numerical simulations averaged over $10^4$ independent realizations of the model. Solid black circles plotted on a black curve express $P(n_s)$ obtained from Eq.~(\ref{defPns4}). Open squares plotted on a gray curve represent the approximated formula (Eq.~(\ref{defPns5})).}
	\label{reffig5}
\end{figure}

Finally, after inserting Eqs.~(\ref{PBell1c}) and~(\ref{PBell2d}) into Eq.~(\ref{defPns3}), one obtains the following exact expression for the probability distribution of the number of clusters $n_s$ of size $s$ in the coagulating system with constant kernel:
\begin{equation}\label{defPns4}
P(n_s)\!=\!\frac{\binom{k}{n_s}}{\binom{N-1}{k-1}} \sum_{\kappa=0}^{\kappa_{max}} \!\binom{k\!-\!n_s}{\kappa}\!\binom{N\!-\!sn_s\!-\!s\kappa\!-\!1} {k\!-\!n_s\!-\!\kappa\!-\!1}(\!-\!1)^\kappa.
\end{equation}

It is easy to show that, in the above expression, in the limit of large $N\!\gg\!1$ and $kN^{-1}\!\ll\!1$, the fraction of successive sum components behaves as $N^{-1}$. This enables one to simplify Eq.~(\ref{defPns4}) by neglecting all terms in the sum except the first one for $\kappa=0$. This way, one obtains a very simple, approximate expression for the distribution $P(n_s)$ in the coagulating systems with a constant kernel, which turns out to be the hypergeometric distribution:
\begin{equation}\label{defPns5}
P(n_s)\!\simeq\!\frac{\binom{k}{n_s}\binom{N-k}{k-n_s}}{\binom{N}{k}}, 
\end{equation}
whose expected value, $\langle n_s\rangle\!=\!k^2N^{-1}$,  for $sn_s\!\ll\!N$, coincides with Eq.~(\ref{nsK1}):
\begin{equation}\label{nsK3}
\langle n_s\rangle\simeq k\frac{\binom{N}{k-2}}{\binom{N}{k-1}} \simeq\frac{k^2}{N}. 
\end{equation} 

Figure~\ref{reffig5} shows that the obtained expression, Eq.~(\ref{defPns4}), perfectly agrees with numerical simulations of the coagulating system with a constant kernel, even for systems that are quite small. In addition, Figure~\ref{reffig5}~b) shows that for $sn_s\ll N$ the difference between the exact and the approximate formula for $P(n_s)$ is almost nonexistent.

\section{Concluding remarks}\label{sectIV}

What is new in this paper? Unlike in most previous approaches, in our approach, time is discrete. We assume that a single coagulation act occurs in each time step, which causes a direct relationship between the total number of clusters, $k$, and the time, $t$. In other words, the probability that at time $t$, there are exactly $k$ clusters in the system, is given by: $P(k,t)=\delta_{k,N-t}$. This assumption does not diminish the generality of our approach, because the appropriate results for the continuous-time coagulation process can easily be obtained from the discrete-time results, provided that the distribution $P(k,t)$ is known. In such a case, instead of using Eq.~(\ref{defPO}) for the probability $P(\Omega)$ that the system can be found in a state $\Omega$, one would have the product, $P(\Omega)P(k,t)$. In addition, let us note that the distribution $P(k,t)$ is usually not difficult to calculate (see for example, Eq.~(16) in Section~4 in \cite{1974_Bayewitz}, where $P(k,t)$ for the constant kernel is given).

Can the results presented here be developed further? Throughout the paper, to illustrate our approach, we have used only the coagulation process with a constant kernel and monodisperse initial conditions. However, it should be noted that the results obtained can be used to describe systems with a constant kernel and arbitrary initial cluster size distribution $P(s_{0})$. From an algorithmic point of view, it is easy to imagine how such a coagulating system could be obtained. It could be done, for example, by replacing monomers in the originally monodisperse cluster configuration with initial clusters of size $s_{0}$ with probability $P(s_0)$. From the point of view of mathematical description, the resulting composite clusters could be analyzed within the random sum formalism \cite{Holst,Kolchin}, which is suited to describing such composite structures (i.e.,~clusters built from other clusters) \cite{bookWilf}. The mentioned analysis would be of great importance, because it could be used to verify the mean-field scaling solutions corresponding to the constant kernel with arbitrary initial conditions, which not long ago were obtained by mathematicians as the solutions to  Smoluchowski's equation \cite{2004_Menon} but which often are unknown to physicists and chemists \cite{2010_bookKrapivsky, 2003_PhysRepLeyvraz}.

Finally, although in this paper we show only that our approach works in the case of the fixed kernel, we must emphasize that the approach can be used to describe systems with arbitrary kernels and, at least, monodisperse initial conditions. The only adjustment needed to make this possible is to calculate the sequence $\{x_g\}$ in which every variable $x_g$ gives the number of ways in which a cluster of size $g$ can be created.

Where else can one use the results? To answer this, we note that one important field of research is related to percolation phenomena in random networks. Although mutual relationships have long been known to exist among the time evolution of classical random graphs, percolation phenomena, and coagulating systems (see, for example \cite{2005_JPhysALushnokov} or Chap.~14.3 in \cite{2010_bookKrapivsky}), recently, interesting problems related to discontinuous and hybrid (mixed-order) percolation transitions in a wild family of cluster merging processes \cite{2010_PRECho, 2016_PRLCho} were launched and are awaiting theoretical description.

\acknowledgments
This work has been supported by the National Science Centre of Poland (Narodowe Centrum Nauki, NCN) under grants no.~2012/05/E/ST2/02300 (A.F. and P.F.) and no.~2015/18/E/ST2/00560 (A.Ch.).

\end{document}